\documentclass[oneside,english]{elsart}
\usepackage[T1]{fontenc}
\usepackage[latin1]{inputenc}
\usepackage{graphicx}
\usepackage{setspace}

\makeatletter

\providecommand{\tabularnewline}{\\}

\usepackage{babel}
\makeatother
\begin{document}
\begin{frontmatter}

\title{Disassembling the Nuclear Matrix Elements of the 
Neutrinoless $\beta\beta$ Decay}

\author{J. Men\'endez and A. Poves}

\address{Departamento de F\'isica Te\'orica and IFT-UAM/CSIC, Universidad~Aut\'onoma~de~Madrid,
E-28049, Madrid, Spain}

\author{E. Caurier and F. Nowacki}

\address{IPHC, IN2P3-CNRS/Universit\'e Louis Pasteur BP 28, F-67037, Strasbourg~Cedex~2, France}

\begin{abstract}
In this article we analyze the nuclear matrix elements (NME) of the
neutrinoless double beta decays ($0\nu\beta\beta$) of the nuclei
$^{48}$Ca, $^{76}$Ge, $^{82}$Se, $^{124}$Sn, $^{128}$Te, $^{130}$Te and $^{136}$Xe
in the framework of the Interacting Shell Model (ISM). We study the
relative value of the different contributions to the NME's, such as higher
order terms in the nuclear current, finite nuclear size effects and
short range correlations, as well as their evolution with the maximum
seniority permitted in the wave functions. We discuss also the build-up of the NME's 
as a function of the distance between the decaying neutrons.
We calculate the decays to $0_{1}^{+}$ final states and find that these decays are at
least 25 times more suppressed with respect to the ground state to ground state transition.
\end{abstract}

\end{frontmatter}

\section{Introduction}

The discovery of neutrino oscillations in recent experiments at
Super-Kamiokande \cite{PhysRevLett.81.1562}, SNO
\cite{PhysRevLett.89.011301} and KamLAND \cite{PhysRevLett.90.021802}
has changed the old conception of neutrinos by proving that they are
massive particles. According to the origin of their mass, neutrinos
can be either Dirac or Majorana particles, the latter case being
particularly interesting since it would imply an extension to the
standard model of electroweak interactions. In this scenario neutrinos are their
own antiparticles and lepton number is not conserved.
Besides, it happens that the best way to detect one of these
violating processes and consequently to establish the Majorana character
of the neutrinos would be the observation of the neutrinoless double beta decay
($0\nu\beta\beta$).

Double beta decay is a very slow weak process. It takes place between
two even-even isobars when the single beta decay is energetically
forbidden or hindered by large spin difference. Two neutrinos beta
decay is a second order weak process ---the reason of its low rate---,
and has been measured in a few nuclei. The $0\nu\beta\beta$ decay
is analog but needs neutrinos to be Majorana particles. With the exception
of one unconfirmed claim \cite{KlapdorKleingrothaus:2001ke,KlapdorKleingrothaus:2004wj},
it has never been observed, and currently there is a number of experiments
either taking place \cite{arnold:182302,arnaboldi:142501,bloxham:025501}
or expected for the near future ---see e.g. ref. \cite{Avignone:2007fu}---
devoted to detect it and to set up firmly the nature of
neutrinos.

Furthermore, $0\nu\beta\beta$ decay is also sensitive to the absolute
scale of neutrino mass, and hence the mass hierarchy ---at present,
only the difference between different mass eigenstates is known. Since
the half-life of the decay is determined, together with the neutrino masses,
by the nuclear matrix element (NME) for this process, the knowledge
of these NME's is essential to predict the most favorable decays and,
once detection is achieved, to settle the neutrino mass scale and
hierarchy.

Two different and complementary methods are mainly used to calculate
NME's for $0\nu\beta\beta$ decays. One is the family of the quasiparticle
random-phase approximation (QRPA). This method has been used by different
groups and a variety of techniques is employed, with results for most
of the possible emitters \cite{Suhonen:1998ck,Rodin:2006yk}. This
work concerns to the alternative, the interacting shell model (ISM) \cite{Caurier:2004gf}. 

In previous ISM works \cite{Retamosa:1995xt,Caurier:prl.77.1954}, the
NME's for the $0\nu\beta\beta$ decay were calculated taking into account
only the dominant terms of the nucleon current. However, in ref. \cite{Simkovic:1999re}
it was noted that the higher order contributions to the current (HOC) are
not negligible and it was claimed that they could reduce up to 20\%-30\%
the final NME's. Subsequently, other QRPA calculations \cite{Kortelainen:2007rh,Kortelainen:2007mn}
have also taken into account these terms, although resulting in a
somewhat smaller correction. These additional nucleon current contributions
have been recently included for the first time in the ISM framework
\cite{Caurier:2007wq}, and it is the aim of this work to look at
them in more detail in order to grasp their relevance for the NME's.
Then we give a comprehensive overview of the present state-of-the-art
ISM and QRPA results by comparing the relative contribution
of the different contributions to the NME's.
In particular, the short range correlations (SRC) are modeled both
by the Jastrow prescription and by the UCOM method \cite{Feldmeier:1997zh}.
The radial dependence of the NME ---first discussed in ref. \cite{Simkovic:2007vu}--- is also
studied within the ISM.

In addition to the usual calculation of $0\nu\beta\beta$ decays to
the ground states of the final nuclei we have also computed the corresponding
NME's for the decays to the first excited $0_{1}^{+}$ states. This comparison
has already been performed for most emitters using QRPA methods
\cite{Suhonen:2000my,Suhonen:2000dr,Suhonen:2002wi,Suhonen:2003gx,Simkovic:2001qf},
concluding that the transitions to excited states are much more suppressed.
However, it is interesting to test these results from the ISM point
of view, and even to improve them, since in refs. \cite{Suhonen:2000my,Suhonen:2000dr,Suhonen:2002wi,Suhonen:2003gx}
the new hadronic current contributions were neglected and in ref.
\cite{Simkovic:2001qf} the short range correlations were not treated
properly (see ref. \cite{Rodin:2007fz}).

\section{\label{sec:theory} Theoretical framework}

The starting point for the $0\nu\beta\beta$ decay is the weak Hamiltonian:

\begin{equation}
H_{W}=\frac{G}{\sqrt{2}}\left(j_{L\mu}J_{L}^{\mu\dagger}\right)+h.c.,\label{eq:H_w}\end{equation}

where $j_{L\mu}$ is the leptonic current, and the hadronic ---nuclear---
counterpart is given in the impulse approximation by:

\begin{eqnarray}
J_{L}^{\mu\dagger} & = & \overline{\Psi}\tau^{+}\left(g_{V}\left(q^{2}\right)\gamma^{\mu}-ig_{M}\left(q^{2}\right)\frac{\sigma^{\mu\nu}}{2M_{p}}\right.\nonumber \\
 &  & \;\;\;\;\;\;\left.-g_{A}\left(q^{2}\right)\gamma^{\mu}\gamma_{5}+g_{P}\left(q^{2}\right)q^{\mu}\gamma_{5}\right)\Psi,
\label{eq:J_mu}\end{eqnarray}

with $q^{\mu}$ the momentum transferred from hadrons to leptons, this
is, $q^{\mu}=p_{neutron}^{\mu}-p_{proton}^{\mu}$.

In the non relativistic case, and discarding energy transfers between
nucleons, we have:

\begin{equation}
J_{L}^{\mu\dagger}\left(\mathbf{x}\right)=\sum_{n=1}^{A}\tau_{n}^{-}\left(g^{\mu0}J^{0}\left(q^{2}\right)+g^{\mu k}J_{n}^{k}\left(q^{2}\right)\right)\delta\left(\mathbf{\mathbf{x}-r_{n}}\right),\label{eq:J_norel}\end{equation}

where:

\begin{eqnarray}
J^{0}\left(q^{2}\right) & = & g_{V}\left(q^{2}\right),\nonumber \\
\mathbf{J}_{n}\left(q^{2}\right) & = & ig_{M}\left(q^{2}\right)\frac{\mathbf{\sigma}_{n}\times\mathbf{q}}{2M_{p}}+g_{A}\left(q^{2}\right)\mathbf{\sigma}_{n}-g_{P}\left(q^{2}\right)\frac{\mathbf{q}\left(\mathbf{q\sigma_{n}}\right)}{2M_{p}}.\label{eq:Jota}\end{eqnarray}

The parametrization of the couplings by the standard dipole form factor
 ---to take into account the finite nuclear size (FNS)---
and the use of the CVC and PCAC hypotheses ---for the magnetic and
 pseudoscalar couplings $g_{M}$ and $g_{P}$--- are those described
in ref. \cite{Simkovic:1999re}. We take as values of the bare
couplings $g_{V}\left(0\right)=1$ and $g_{A}\left(0\right)=1.25$.

Due to the high momentum of the virtual neutrino in the nucleus ---$\approx$100
MeV--- we can replace the intermediate state energy by an average
value and then use the closure relation to sum over all the intermediate
states. This approximation is correct to better than 90\% \cite{Muto:1994hi}. 

We also limit our study to transitions to $0^{+}$ final states, and
assume electrons to be emitted in $s$ wave. Corrections to these
approximations are of the order of 1\% at most, due to the fact that
in the other cases effective nuclear operators of higher orders are needed
to couple the initial and final states.

With these considerations, the expression for the half-life of the
$0\nu\beta\beta$ decay can be written as \cite{Doi:1980ze,Doi:1985dx}:

\begin{equation}
\left(T_{1/2}^{0\nu\beta\beta}\left(0^{+}\rightarrow0^{+}\right)\right)^{-1}=G_{01}\left|M^{0\nu\beta\beta}\right|^{2}\left(\frac{\left\langle m_{\nu}\right\rangle }{m_{e}}\right)^{2},\label{eq:t-1}\end{equation}

where $\left\langle m_{\nu}\right\rangle =|\sum_{k}U_{ek}^{2}m_{k}|$
is the effective Majorana neutrino mass, a combination of the neutrino masses $m_{k}$ 
due to the neutrino mixing matrix $U$ and $G_{01}$
is a kinematic factor ---dependent on the charge, mass and available
energy of the process. $M^{0\nu\beta\beta}$ is the NME object of
study in this work. As we see, the neutrino
mass scale is directly related to the decay rate.

The kinematic factor $G_{01}$ depends on the value of the coupling constant $g_{A}$. Therefore
the NME's obtained with different $g_{A}$ values cannot be directly compared.
If we redefine the NME as:

\begin{equation}
M'^{\;0\nu\beta\beta}=\left(\frac{g_{A}}{1.25}\right)^{2}M^{0\nu\beta\beta}.\label{eq:mprime}
\end{equation}

These $M'^{\;0\nu\beta\beta}$'s are directly comparable no matter
which was the value of $g_{A}$ employed in their calculation, since they share a common
$G_{01}$ factor ---that of $g_{A}=1.25$. In this sense, the translation of
$M'^{\;0\nu\beta\beta}$'s into half-lives is transparent. The QRPA results obtained with different
$g_{A}$ values are already expressed in this way by the authors of refs. \cite{Simkovic:2007vu,Rodin:2007fz}
while the results of refs. \cite{Kortelainen:2007rh,Kortelainen:2007mn,Suhonen:2008ijmpa} are not. The latter have been translated
into the form of eq. (\ref{eq:mprime}) when comparing to our results.

Looking back at the NME, it is obtained from the effective transition
operator resulting of the product of the nuclear currents:

\begin{equation}
\Omega\left(q\right)=-h^{F}\left(q\right)+h^{GT}\left(q\right)\mathbf{\mathbf{\sigma}_{n}\mathbf{\sigma}_{m}}-h^{T}\left(q\right)\mathbf{S}_{nm}^{q},\label{eq:sigma_q}\end{equation}

where $\mathbf{S}_{nm}^{q}=3\left(\mathbf{\hat{q}\sigma_{n}\hat{q}\sigma_{m}}\right)-\mathbf{\mathbf{\sigma}_{n}\mathbf{\sigma}_{m}}$
is the tensor operator. The functions $h\left(q\right)$ can be labeled
according to the current terms from which they come from:

\begin{eqnarray}
h^{F}\left(q\right) & = & h_{vv}^{F}\left(q\right),\nonumber \\
h^{GT}\left(q\right) & = & h_{aa}^{GT}\left(q\right)+h_{ap}^{GT}\left(q\right)+h_{pp}^{GT}\left(q\right)+h_{mm}^{GT}\left(q\right),\nonumber \\
h^{T}\left(q\right) & = & h_{ap}^{T}\left(q\right)+h_{pp}^{T}\left(q\right)+h_{mm}^{T}\left(q\right),\label{eq:h_xx}\end{eqnarray}

whose explicit form can be found in ref. \cite{Simkovic:1999re}.

In ISM works previous to ref. \cite{Caurier:2007wq}, only $h_{aa}$ and
$h_{vv}$ terms were considered. However, rough estimates of the value
of all terms taking $q\approx100$ MeV give: $h_{aa}\approx h_{vv}\approx1$,
$h_{ap}\approx0.20$, $h_{pp}\approx0.04$ and $h_{mm}\approx0.02$.
Therefore, according to this figures, certainly $h_{ap}$ cannot
be neglected. Since the Gamow-Teller contribution will be the dominant
one, and both the $h_{pp}$ and $h_{mm}$ have the
same sign and opposite to $h_{ap}$, it seems sensible to keep all
these terms in the calculation. In Section \ref{sec:Results-and-discussion}
we will see how these terms indeed contribute to the total NME.

Integrating over $\mathbf{q}$ we get the corresponding operators
in position space, which are called the neutrino potentials. Before
radial integration they look like:

\begin{eqnarray}
V_{x}^{F/GT}\left(r\right) & = & \frac{2}{\pi}\frac{R}{g_{A}^{2}\left(0\right)}\int_{0}^{\infty}j_{0}\left(qr\right)\frac{h_{x}^{F/GT}\left(q\right)}{\left(q+\mu\right)}q\, dq,\nonumber \\
V_{x}^{T}\left(r\right) & = & \frac{2}{\pi}\frac{R}{g_{A}^{2}\left(0\right)}\int_{0}^{\infty}-j_{2}\left(qr\right)\frac{h_{x}^{T}\left(q\right)}{\left(q+\mu\right)}q\, dq,\label{eq:V_int}\end{eqnarray}

where $j_{n}\left(x\right)$ are the spherical Bessel functions, $r$
is the distance between nucleons and $R$, which makes the result
dimensionless, is taken as $R$=$r_{0}A^{1/3}$, with $r_{0}=1.2$
fm.

Finally, the NME reads:

\newpage

\begin{eqnarray}
M^{0\nu\beta\beta} & = & -\left(\frac{g_{V}\left(0\right)}{g_{A}\left(0\right)}\right)^{2}M^{F}+M^{GT}-M^{T}\nonumber \\
 & = & \left\langle 0_{f}^{+}\right|\sum_{n,m}\tau_{n}^{-}\tau_{m}^{-}\left(-V^{F}\left(r\right)+V^{GT}\left(r\right)\mathbf{\mathbf{\sigma}_{n}\mathbf{\sigma}_{m}}-V^{T}\left(r\right)\mathbf{\mathbf{S}_{nm}^{r}}\right)\left|0_{i}^{+}\right\rangle .\label{eq:NME}\end{eqnarray}

In the calculation of the NME we need to take into account the
short range correlations. Using a standard Jastrow type
function, the NME's are modified as \cite{Wu:1985xy,Miller:1975hu}:

\begin{eqnarray} \langle 0_{f}^{+}|V\left(r\right)|0_{i}^{+}\rangle _{src} & = & \langle 0_{f}^{+}f\left(r\right)|V\left(r\right)|f\left(r\right)0_{i}^{+}\rangle \nonumber \\  & = & \langle 0_{f}^{+}|f\left(r\right)^{2}V\left(r\right)|0_{i}^{+}\rangle, \label{eq:src_me}\end{eqnarray}

with $f\left(r\right)=1-e^{-ar^{2}}\left(1-br^{2}\right)$, where
$a=1.1\textrm{ fm}^{-2}$ and $b=0.68\textrm{ fm}^{-2}$.

\section{\label{sec:Results-and-discussion}The NME's disassembled}

Following these considerations, we have performed calculations for
the $0\nu\beta\beta$ decays of the emitters $^{48}$Ca, $^{76}$Ge,
$^{82}$Se, $^{124}$Sn, $^{128}$Te, $^{130}$Te and $^{136}$Xe, using the ISM
coupled code described in ref. \cite{Caurier:2004gf}, ideally adapted for
the calculation of $0^{+}$
states. Full diagonalizations are accomplished within different valence spaces
and effective interactions. For instance, the decay of $^{48}$Ca
is studied in the $pf$ major shell, and the KB3 interaction is employed.
For the case of $^{76}$Ge and $^{82}$Se, the valence space consisting
on 1p$_{3/2}$, 0f$_{5/2}$, 1p$_{1/2}$ and 0g$_{9/2}$ is diagonalized
using the GCN28.50 interaction. Finally the 0g$_{7/2}$, 1d$_{3/2}$,
1d$_{5/2}$, 2s$_{1/2}$ and 0h$_{11/2}$ valence space and the GCN50.82
interaction are used in the decays of $^{124}$Sn, $^{128}$Te, $^{130}$Te and
$^{136}$Xe. The latter spaces and interactions will be discussed
in detail elsewhere \cite{Gniady:0000aa}.

QRPA valence spaces comprise at least two major oscillator shells.
The effect of the orbits excluded in our ISM calculations was studied in
Ref.~\cite{Caurier:2007qn}, in the particular cases of $A=82$ and $A=136$.
Proton and neutron 2p-2h excitations were considered separately,
obtaining an increase in the NME's not larger than 20\%. However, this
number may be an overestimation since the occupancies obtained
for the extra orbits are larger than that of QRPA calculations.
Furthermore, when more than one orbit was added to the valence space,
the NME increased in the same amount as with only one extra orbit,
showing that different contributions do not sum up.
An extended study of these effects is currently in preparation.

The results of the Shell Model calculations were shown in ref.~\cite{Caurier:2007wq},
where a comparison of the final values to the ones from recent
QRPA calculations was made, pointing out the importance of including
all the seniority components in a given space.
Here we will have
a closer look at the NME's, in order to gain insight into the details
of the calculation and recognise possible uncertainties.

\begin{table}
\caption{\label{cap:hoc_evol}Evolution of the $^{76}$Ge~$\rightarrow$~$^{76}$Se
NME's described in Section \ref{sec:theory} as we successively add
the $ap$, $pp$ and $mm$ HOC contributions.}
\begin{center}
\smallskip
\begin{tabular*}{\linewidth}{@{\extracolsep{\fill}}c|cccc}
\hline \noalign{\smallskip}

&
$vv+aa$&
$+ap$&
$+pp$&
$+mm$\tabularnewline
\hline
$M_{hoc}$&
4.04&
2.82&
3.29&
3.29\tabularnewline
$M_{hoc+fns}$&
3.45&
2.49&
2.80&
2.96\tabularnewline
$M_{hoc+src}$&
2.85&
2.12&
2.36&
2.36\tabularnewline
$M^{0\nu\beta\beta}$&
2.70&
2.01&
2.21&
2.30\tabularnewline
$\Delta M^{0\nu\beta\beta}$&
&
-25\%&
+7\%&
+3\%\tabularnewline
\hline
\end{tabular*}\end{center}
\end{table}

In Table \ref{cap:hoc_evol} we show the contribution of the additional
nucleon current terms in the particular case of $^{76}$Ge~$\rightarrow$~$^{76}$Se
$0\nu\beta\beta$ decay. In addition to the final result we also give the
partial ones not considering FNS and/or SRC. We see that these terms
contribute to reduce the bare matrix element in approximately 20\%.
However, as was pointed out in ref. \cite{Caurier:2007wq}, the effect
of the HOC contributions in the full NME ---FNS and SRC taken into account---
is a bit smaller, 15\%, which only means that HOC contributions are
slightly more regularized by FNS and SRC than the lower order ---$aa$
and $vv$--- terms. In addition to that, the $mm$ contribution to
HOC, which increases the NME and hence reduces the net HOC effect,
vanishes in the bare case. The behaviour of all other decays is very
similar. From this table we also conclude that the overall HOC contribution
is of the expected relative value, and, moreover, the individual terms
are also very close to their estimate in Section \ref{sec:theory}; they
are found only a bit larger, but with the correct relative figures.

It has recently been discussed in ref.
\cite{Caurier:2007wq} that the pairing interaction favours the $0\nu\beta\beta$
decay and that, consequently, truncations in seniority tend
to overestimate the value of the NME's. This is,
when we perform truncations in seniority in the initial and final wave functions,
the value of the full NME is not converged, in general, until the full seniority calculation
is done. This may explain the difference of the results ---due to their different treatment 
of pairing type correlations--- for the NME's obtained between the ISM and QRPA approaches.

High seniority components are strongly connected to nuclear deformation.
As an example, we show in Table~\ref{cap:66_senwf} the decomposition
of the wave function of the nucleus $^{66}$Ge
---that would exhibit a ficticious $0\nu\beta\beta$ decay to its mirror $^{66}$Se---
for different deformations,
obtained by adding a variable extra quadrupole-quadrupole term to the interaction.
We see that, as the nucleus becomes more deformed, the high seniority components
become more important.

\begin{table}
\caption{Decomposition of the wave function of the ground state of $^{66}$Ge
according to its seniority components, in percentage, for different values
of the deformation $\beta$.
\label{cap:66_senwf}}
\begin{center}
\smallskip
\begin{tabular*}{\linewidth}{@{\extracolsep{\fill}}c|ccccc}
\hline \noalign{\smallskip}
$\beta$ & $s=0$ & $s=4$ & $s=6$ & $s=8$ & $s=10$ \\
\hline
0.15 & 78 & 20 &  1 &  1 &  0 \\
0.20 & 39 & 43 &  7 & 10 &  1 \\
0.25 & 20 & 43 & 14 & 20 &  3 \\
0.30 &  6 & 32 & 21 & 31 & 10\\
\hline
\end{tabular*}\end{center}
\end{table}

We can check if the discrepancy of the NME's remains in the HOC, FNS and SRC contributions.
For that purpose, in Table \ref{cap:nme_comp} we compare in detail the NME's for $A=76$
for ISM and QRPA calculations, enclosing the partial results as well.
We also append the amount of correction due to each HOC, FNS and SRC
to the previous step in the calculation. The SRC are always of the
Jastrow type unless otherwise stated; we will return to this point
later. Similar figures are found for any other decay. If we focus
in the relative importance of the different approximations, we can
see that all three calculations show the same trend, though the details
may change a bit from one to another. Hence we can conclude that HOC
contributions reduce the NME about 10-20\%, taking into account
FNS effects produces an additional 10-20\% decrease and finally Jastrow
type SRC reduce the NME 20-25\%. The overall effect
from the original bare NME adds up to 35-45\%. In our calculations,
these three contributions when applied to the bare NME, give reductions of
20\%, 15\% and 30\%, respectively ---$M_{src}=2.85$, not included in
Table \ref{cap:nme_comp}. Therefore, these effects do not pile up, the total
reduction amounting to roughly 70\% of the sum of the individual contributions.

\begin{table}
\caption{\label{cap:nme_comp} Comparison of the NME for the $^{76}$Ge $\rightarrow$
$^{76}$Se $0\nu\beta\beta$ decay for this work (ISM) and the Jyv\"askyl\"a
(JY07) and T\"ubingen (TU99) groups. The values of TU99 were originally
calculated with $r_{0}=1.1$ fm, and have been corrected to be directly
comparable with the others. }

\begin{center}
\smallskip
\begin{tabular*}{\linewidth}{@{\extracolsep{\fill}}c|cccccccc}
\hline \noalign{\smallskip}
&
$M_{bare}$&
$M_{fns}$&
$M_{hoc}$&
$M_{h.+fns}$&
$M^{0\nu\beta\beta}$&
\%$_{hoc}^{b/fns}$&
\%$_{fns}$&
\%$_{src}$\tabularnewline
\hline
ISM&
4.04&
3.45&
3.29&
2.96&
2.30&
$19/14$&
10&
22\tabularnewline
JY07 \cite{Kortelainen:2007rh}&
8.53&
-&
7.72&
6.36&
4.72&
$9/-$&
18&
26\tabularnewline
TU99 \cite{Simkovic:1999re}&
-&
7.03&
-&
5.63&
-&
$-/20$&
$\approx10$&
$\approx20$\tabularnewline
\hline
\end{tabular*}\end{center}
\end{table}

\begin{table}
\caption{\label{cap:senior_76ge}Evolution of the NME with the maximum seniority  ($s_{m}$)
permitted in the wave functions of  $^{76}$Ge and 
$^{76}$Se, including different contributions to the full operator.}

\begin{center}
\smallskip
\begin{tabular*}{\linewidth}{@{\extracolsep{\fill}}c|cccccccc}
\hline \noalign{\smallskip}

$s_{m}$&
$M_{bare}$&
$M_{fns}$&
$M_{hoc}$&
$M_{h.+fns}$&
$M^{0\nu\beta\beta}$&
\%$_{hoc}^{b/fns}$&
\%$_{fns}$&
\%$_{src}$\tabularnewline
\hline
0&
12.31&
11.16&
10.49&
9.83&
8.59&
15/12&
6&
13\tabularnewline
4&
8.84&
7.87&
7.44&
6.89&
5.82&
16/12&
7&
16\tabularnewline
6&
8.01&
7.11&
6.73&
6.22&
5.23&
16/13&
8&
16\tabularnewline
8&
5.63&
4.90&
4.66&
4.25&
3.34&
17/13&
9&
21\tabularnewline
10&
4.64&
4.00&
3.81&
3.45&
2.74&
18/14&
9&
21\tabularnewline
12&
4.10&
3.50&
3.34&
3.01&
2.34&
19/14&
10&
22\tabularnewline
14&
4.04&
3.45&
3.29&
2.96&
2.30&
19/14&
10&
22\tabularnewline
\hline
\end{tabular*}\end{center}
\end{table}

Therefore, contrary to what happens for the NME's, we see that
the relative contributions of HOC, FNS or SRC are similar for ISM and QRPA calculations,
which seems to point out that their contribution to the full NME is not affected by seniority truncations.
This is confirmed in Table \ref{cap:senior_76ge}, where
these partial contributions as a function of the seniority are shown. The $^{76}$Ge~$\rightarrow$~$^{76}$Se
$0\nu\beta\beta$ decay is chosen again, but the same conclusion is
obtained for all other transitions. Even though a small decrease of all the contributions
is seen, at the $s_{m}=4$ level---leading order in QRPA--- the relative value of
HOC, FNS and SRC is essentially that of the full calculation.

\begin{table}
\caption{\label{cap:senior_xif}Evolution of $-\chi^{F}$
as a function of the maximum seniority allowed in the wave functions, $s_{m}$, for all
the studied $0\nu\beta\beta$ decays.}

\begin{center}
\smallskip
\begin{tabular*}{\linewidth}{@{\extracolsep{\fill}}c|ccccccc}
\hline \noalign{\smallskip}

$s_{m}$&
$A=48$&
$A=76$&
$A=82$&
$A=124$&
$A=128$&
$A=130$&
$A=136$\tabularnewline
\hline
0&
0.33&
0.31&
0.30&
0.27&
0.27&
0.27&
0.26\tabularnewline
4&
0.19&
0.23&
0.21&
0.15&
0.20&
0.19&
0.15\tabularnewline
6&
0.17&
0.22&
0.20&
0.16&
0.20&
0.19&
0.16\tabularnewline
8&
0.16&
0.16&
0.13&
0.15&
0.16&
0.15&
0.15\tabularnewline
10&
&
0.14&
0.12&
0.15&
0.15&
0.15&
\tabularnewline
12&
&
0.12&
0.11&
0.15&
0.15&
&
\tabularnewline
14&
&
0.12&
&
&
&
\tabularnewline
\hline
\end{tabular*}\end{center}
\end{table}

Another aspect of the NME that could be sensitive to the treatment of the pairing correlations
is the ratio of Fermi to Gamow-Teller terms by means of the coefficient  $\chi^{F}=\left(\frac{g_{V}\left(0\right)}{g_{A}\left(0\right)}\right)^{2}M^{F}/M^{GT}$, 
represented in Table \ref{cap:senior_xif}
for all the studied nuclei as a function of the seniority. Unlike
the precedent case, here we see that correlations affect in
a different manner to these contributions, in such a way that the
ratio $\chi^{F}$ decreases as we allow higher seniority components in the wave
functions. This trend is not seen, however, in the $A=48$, $A=124$
and $A=136$ cases. But these are precisely the nuclei for which the
low seniority truncation works better \cite{Caurier:2007wq}, being
emitters which, in their natural valence spaces, only consist on neutrons
($^{48}$Ca and $^{124}$Sn) or protons ($^{136}$Xe), leading to wave
functions dominated by low seniority components. We compare our full
results and those truncated in seniority with the QRPA figures in Table
\ref{cap:compare_xi}, observing that ISM $\chi^{F}$ values are smaller
than QRPA's, but the truncated $s_{m}=4$ results are always closer
to them. This is to say, one may attribute the discrepancy in $\chi^{F}$
between ISM and QRPA to the seniority truncations, as was the
case for the complete NME.

\begin{table}
\caption{\label{cap:compare_xi}Comparison of the values of $\chi^{F}$ of
this work (ISM), this work with seniority 
$s_ m=4$ and the QRPA results of the Jyv\"askyl\"a (JY07) and T\"ubingen (TU07) groups. The
TU07 result is taken prior to SRC correlations, see ref. \protect\cite{Rodin:2007fz}.}

\begin{center}
\smallskip
\begin{tabular*}{\linewidth}{@{\extracolsep{\fill}}cc|cccccc}
\hline \noalign{\smallskip}

\multicolumn{2}{c|}{$-\chi^{F}$}&
&
ISM&
&
ISM $\left(s_m=4\right)$&
JY07 \cite{Kortelainen:2007rh,Kortelainen:2007mn}&
TU07 \cite{Simkovic:1999re}\tabularnewline
\hline
\multicolumn{2}{c|}{$^{76}$Ge $\rightarrow$ $^{76}$Se}&
&
0.12&
&
0.23&
0.27&
0.32\tabularnewline
\multicolumn{2}{c|}{$^{82}$Se $\rightarrow$ $^{82}$Kr}&
&
0.11&
&
0.21&
0.26&
-\tabularnewline
\multicolumn{2}{c|}{$^{128}$Te $\rightarrow$ $^{130}$Xe}&
&
0.15&
&
0.20&
0.31&
-\tabularnewline
\multicolumn{2}{c|}{$^{130}$Te $\rightarrow$ $^{130}$Xe}&
&
0.15&
&
0.19&
0.31&
0.36\tabularnewline
\multicolumn{2}{c|}{$^{136}$Xe $\rightarrow$ $^{136}$Ba}&
&
0.15&
&
0.15&
0.27&
-\tabularnewline
\hline
\end{tabular*}\end{center}
\end{table}

So far we have argued that there is agreement in the relative importance of 
the different pieces of the $0\nu\beta\beta$
NME studied, i.e. HOC, FNS and SRC ---of Jastrow
type--- between the ISM and QRPA in its different
versions, while the difference in $\chi^{F}$ may be attributed, 
as in the case of the full NME, to the fact that high seniority components
may not be fully included within the QRPA approach.
However, there are still two issues regarding which there are discrepancies
between the different methods and authors.

On one hand there is the role of tensor part of the NME, which we
can quantify similarly to the Fermi case by the ratio to the Gamow-Teller
contribution, $\chi^{T}=M^{T}/M^{GT}$. While for some QRPA authors \cite{Simkovic:1999re}
this quantity amounts to 5\% for $A=76$ and 8\% for $A=130$ ---prior
to SRC, after that their influence is necessarily larger since tensor
type contributions, contrary to Fermi and Gamow-Teller parts, are not
reduced by SRC---, others claim that their contribution is negligible
\cite{Kortelainen:2007rh}. Our ISM results are shown in Table \ref{cap:senior_xi_t},
collected again as a function of the seniority to explore a possible
dependence of $\chi^{T}$. We can see that, except in
the special case of $A=48$, which shows a quite large ratio, all the
other numbers are far from these found in ref. \cite{Simkovic:1999re},
and could be considered negligible as in ref. \cite{Kortelainen:2007rh}.
This remains true although to a slightly lesser extent if we keep
only the $s_{m}=4$ figures, which are larger than the corresponding
complete space values ---thus pointing to a similar dependence on
the seniority truncation to that of the Fermi component--- but still very minor to
be significant in the final result.

\begin{table}
\caption{\label{cap:senior_xi_t}Evolution of $\chi^{T}$ (\%)
as a function of the maximum seniority permitted, $s_{m}$, in the
wave functions for all the studied $0\nu\beta\beta$ decays.}

\begin{center}
\smallskip
\begin{tabular*}{\linewidth}{@{\extracolsep{\fill}}c|ccccccc}
\hline \noalign{\smallskip}

$s_{m}$&
$A=48$&
$A=76$&
$A=82$&
$A=124$&
$A=128$&
$A=130$&
$A=136$\tabularnewline
\hline
0&
3.2&
1.6&
1.3&
0.5&
0.6&
0.5&
0.4\tabularnewline
4&
9.0&
1.0&
0.8&
0.1&
0.0&
-0.2&
-0.7\tabularnewline
6&
9.4&
0.9&
0.9&
0.1&
0.1&
-0.1&
-0.6\tabularnewline
8&
9.9&
0.6&
0.3&
0.1&
-0.3&
-0.5&
-0.6\tabularnewline
10&
&
0.6&
0.2&
0.1&
-0.3&
-0.5&
\tabularnewline
12&
&
0.4&
0.2&
0.1&
-0.3&
&
\tabularnewline
14&
&
0.4&
&
&
&
&
\tabularnewline
\hline
\end{tabular*}\end{center}
\end{table}

On the other hand, a fully consistent treatment of the SRC's would demand regularizing the $0\nu\beta\beta$
operator using the same prescription than for the bare interaction.
However, this approach is beyond present ISM or QRPA capabilities. Hence, general prescriptions,
that also come from the regularization of bare interactions into the nuclear medium,
are used instead.

The results presented so far have been obtained using the standard Jastrow type correlator of eq. (\ref{eq:src_me}).
Other authors \cite{Kortelainen:2007rn} have recently argued that this correction is somewhat too aggressive,
and have proposed another method ---namely the Unitary Correlation
Operator Method (UCOM) \cite{Feldmeier:1997zh}--- to estimate the SRC, which leads
to a much smoother correction, of the order of 5\% compared to
the 15-25\% of the Jastrow correlator, see Table \ref{cap:nme_comp}. 

We have estimated the value of our ISM results taking an UCOM type
SRC by simulating the correlator as that of the $ST=01$ channel \cite{Roth:2005pd}, 
common throughout the calculation. The correlator of the
other important ---even--- channel is very similar to this one, and the difference should not
change our estimated results. The numbers obtained are
listed in Table \ref{cap:src}, where we find a 5\% reduction
of the NME with this UCOM SRC ansatz, in agreement with the QRPA calculations.
Thus, treating the SRC with a softer prescription of this type increases
our Jastrow correlated final results by some 20\%, leaving the full
reduction of the NME due to all contributions at around 25-35\%.

\begin{figure}

\begin{center}\includegraphics[%
  scale=0.55,
  angle=270]{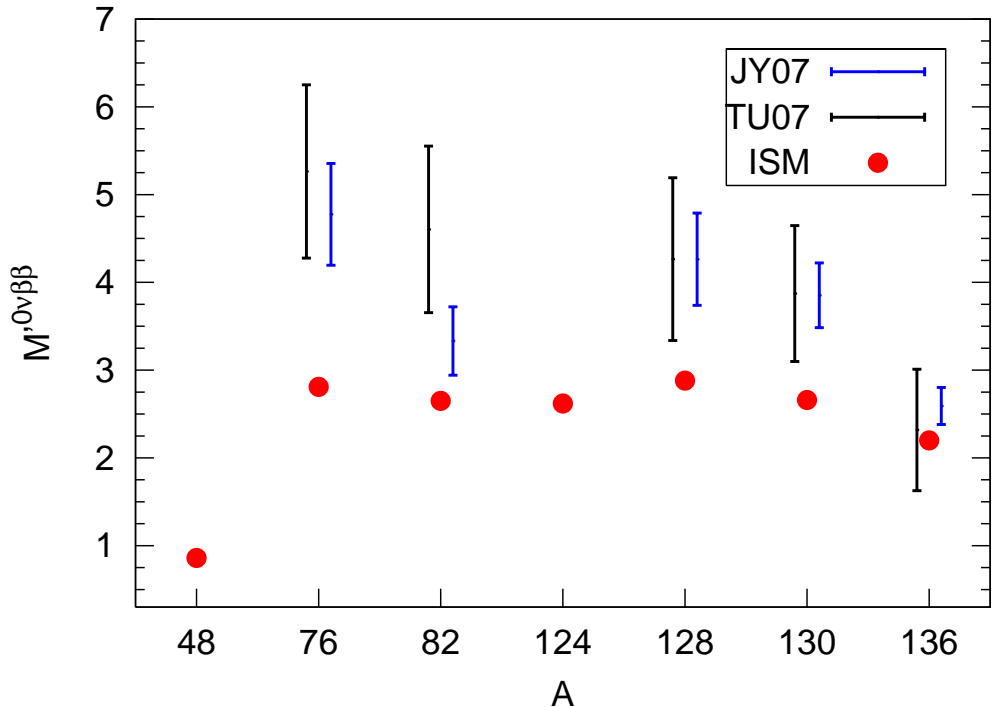}\end{center}

\caption{\label{cap:m0nbb_ucom} The neutrinoless double beta decay $M'^{\;0\nu\beta\beta}$'s for ISM and QRPA 
calculations treating the SRC with the UCOM approach. Tu07 QRPA results from
ref. \protect\cite{Simkovic:2007vu} and Jy07 results from refs. \protect\cite{Kortelainen:2007rh,Kortelainen:2007mn}.}

\end{figure}

\begin{figure}

\begin{center}\includegraphics[%
  scale=0.55,
  angle=270]{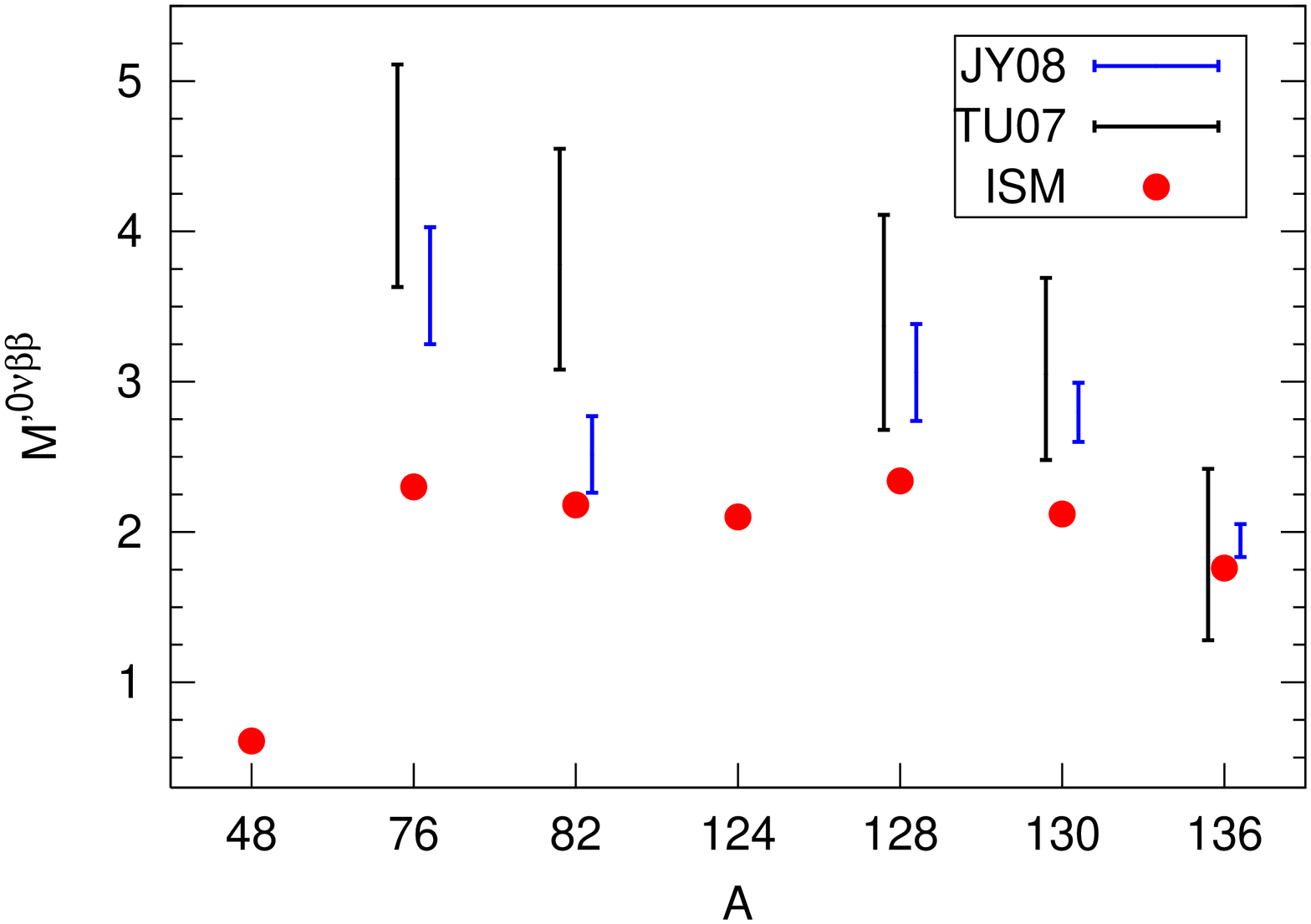}\end{center}

\caption{\label{cap:m0nbb} Same as fig.\ref{cap:m0nbb_ucom} but with Jastrow type SRC. Tu07 QRPA results from 
ref. \protect\cite{Rodin:2007fz} and Jy08 results from ref. \protect\cite{Suhonen:2008ijmpa}.}

\end{figure}

In figure \ref{cap:m0nbb_ucom} the ISM and QRPA results for the NME's 
are compared within this UCOM treatment of the SRC. The same figure but considering
Jastrow type SRC was shown in ref. \cite{Caurier:2007wq}, but inadvertently the results obtained
with different values of $g_A$ where not rescaled properly. This is corrected in figure \ref{cap:m0nbb},
where the comparable $M'^{\;0\nu\beta\beta}$'s introduced in eq. (\ref{eq:mprime}) are represented.
By comparing both figures, it is confirmed that
there is a common trend; the QRPA values are larger than the ISM ones when the high seniority components
are important in the latter. For both ISM and QRPA 
the only net effect of UCOM is an enhancement of the Jastrow results of about 20\%.

Whether the UCOM or Jastrow method is more appropriate to treat the $0\nu\beta\beta$
short range correlations is still an open question. Therefore, taking into account the limitations of our method regarding the SRC's, this different results obtained by both
prescriptions may be considered as an estimation of the range of the effect of SRC's ---5-20\%.

\begin{table}
\caption{\label{cap:src}Contributions of the Jastrow and 
UCOM type SRC to the NME. Jastrow results of ref.\protect\cite{Caurier:2007wq} were slightly
different from present because only one cutoff parameter was considered in the FNS terms.}

\begin{center}
\smallskip
\begin{tabular*}{\linewidth}{@{\extracolsep{\fill}}c|ccccc}
\hline \noalign{\smallskip}

$0\nu\beta\beta$ Transition&
$M_{no\: SRC}$&
$M_{UCOM}^{0\nu\beta\beta}$&
$M_{Jastrow}^{0\nu\beta\beta}$&
$\Delta M_{UCOM}$(\%)&
$\Delta M_{Jastrow}$(\%)\tabularnewline
\hline
$^{48}$Ca $\rightarrow$ $^{48}$Ti&
0.92&
0.85&
0.61&
8&
34\tabularnewline
$^{76}$Ge $\rightarrow$ $^{76}$Se&
2.96&
2.81&
2.30&
5&
22\tabularnewline
$^{82}$Se $\rightarrow$ $^{82}$Kr&
2.79&
2.64&
2.18&
5&
22\tabularnewline
$^{124}$Sn$\rightarrow$ $^{124}$Te&
2.77&
2.62&
2.10&
5&
24\tabularnewline
$^{128}$Te $\rightarrow$ $^{128}$Xe&
3.05&
2.88&
2.34&
6&
23\tabularnewline
$^{130}$Te $\rightarrow$ $^{130}$Xe&
2.81&
2.65&
2.12&
6&
25\tabularnewline
$^{136}$Xe $\rightarrow$ $^{136}$Ba&
2.32&
2.19&
1.76&
6&
24\tabularnewline
\hline
\end{tabular*}\end{center}

\end{table}

\section{Radial dependence of the NME}

Very recently the radial evolution of the NME has been studied in ref. 
\cite{Simkovic:2007vu}, in order to see for what internucleonic distances
$r$ the NME gets the major contribution. This is done by representing
the operator $C(r)$ defined as:\begin{eqnarray}
M^{0\nu\beta\beta} & = & \int_{0}^{\infty}C(r)dr\label{eq:C(r)}\end{eqnarray}

\begin{figure}

\begin{center}\includegraphics[%
  scale=0.55,
  angle=270]{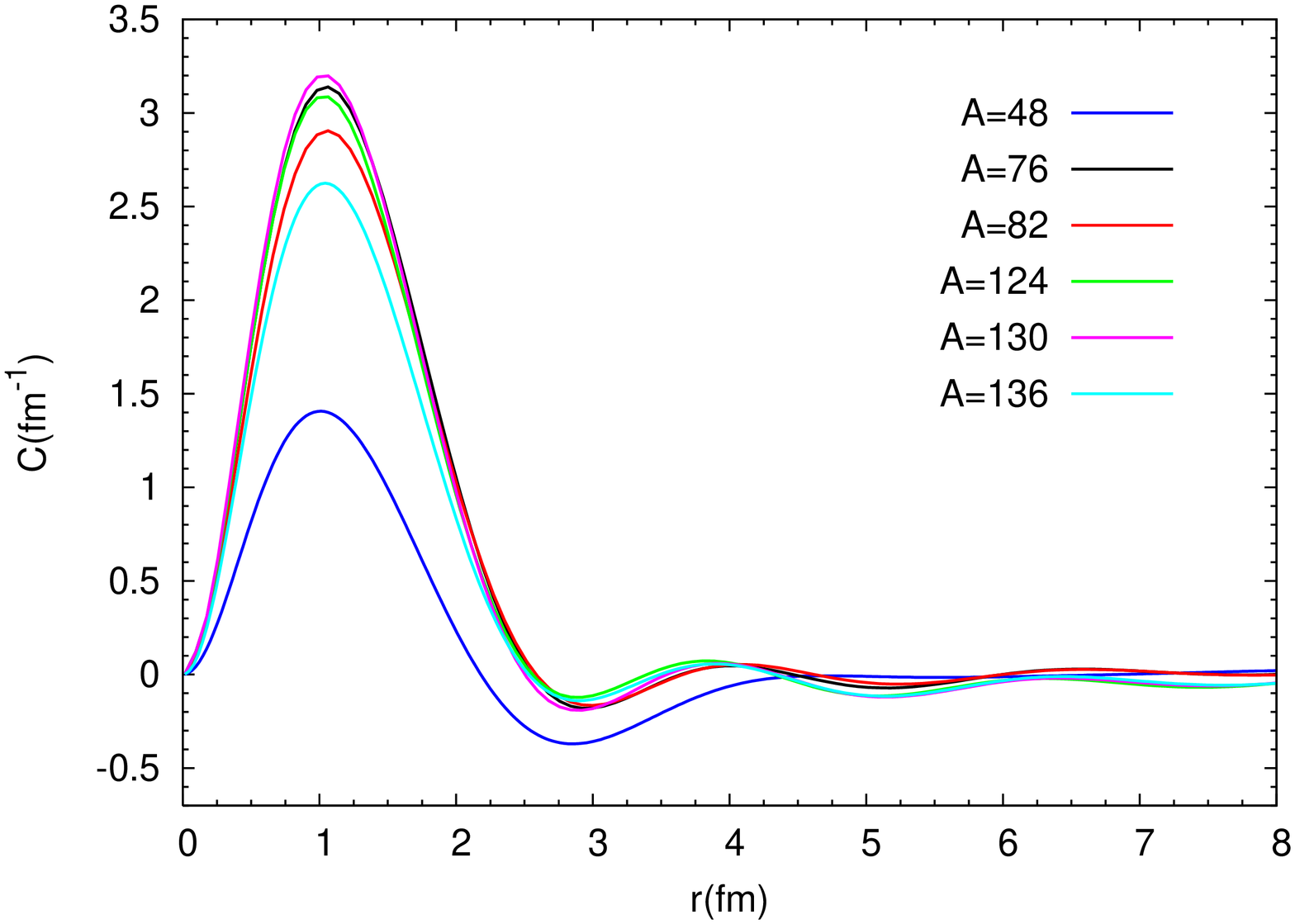}\end{center}

\caption{\label{f:A_comp} Evolution of $C(r)$ for the $^{48}$Ca~$\rightarrow$~$^{48}$Ti,
$^{82}$Se~$\rightarrow$~$^{82}$Kr, $^{124}$Sn~$\rightarrow$~$^{124}$Te,
$^{130}$Te~$\rightarrow$~$^{130}$Xe and $^{136}$Xe~$\rightarrow$~$^{136}$Ba
transitions. SRC are not included in the calculation.}

\end{figure}

\begin{figure}

\begin{center}\includegraphics[%
  scale=0.55,
  angle=270]{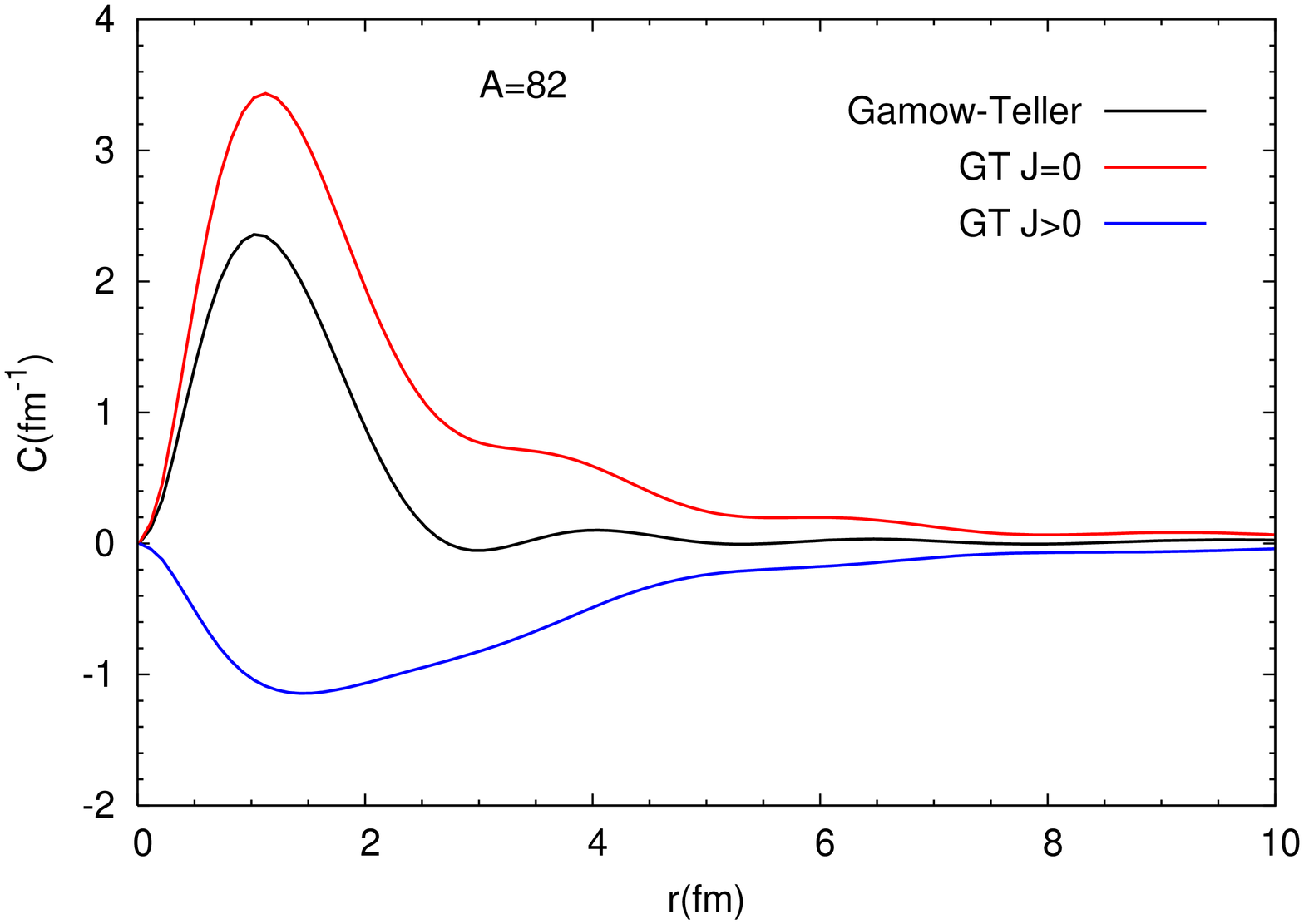}\end{center}

\caption{\label{f:82_jdec} Evolution of the GT part of $C(r)$ for the $^{82}$Se~$\rightarrow$~$^{82}$Kr
transition. The contribution of decaying pairs coupled to $J=0$ and
$J>0$ is also shown. SRC are not included.}

\end{figure}

The form of this operator for the transitions studied is shown in
Figure~\ref{f:A_comp}. This result is in full agreement with the QRPA, confirming the
findings of ref. \cite{Simkovic:2007vu}. This is,  beyond $r=3\, fm$ there is no overall contribution
to the NME, while the maximum value of $C(r)$ occurs around $r=1\, fm$, which means that
almost the complete value of the NME comes from the contribution of decaying nucleons which are
close to each other. This distance corresponds to a momentum of $q\approx$ 200 MeV,
twice the expected value estimated in Section \ref{sec:theory}.
Such a small distance is partly due to the cancellation that
happens between the contribution of decaying pairs coupled to $J=0$
and $J>0$, as can be seen for the dominant GT component of the $^{82}$Se~$\rightarrow$~$^{82}$Kr
$0\nu\beta\beta$ decay in Fig.~\ref{f:82_jdec} ---for all the other transitions the
same tendency is reproduced.
That the NME's radial shape of the QRPA and ISM calculations be identical is quite intriguing and 
perhaps points to a hidden simplicity in their physics. In fact, one can argue
that QRPA and ISM NME's, radial dependence included, differ only by a scaling factor, which 
can be expressed as the ratio of the average number of pairs in both calculations.

\section{Decays to $0_{1}^{+}$ excited states}

When considering the $0\nu\beta\beta$ process, the most favourable decay
channel is, due to phase space considerations, the $0^{+}$ ground state to
ground state one, which is the case of all transitions so
far presented in this paper. However, for experimental reasons, it
would be very interesting to have a decay to an excited state if it is not
too much suppressed compared to the decay to the $0_{g.s.}^{+}$, because
the background reduction, coming from the simultaneous detection
of the electrons from the $0\nu\beta\beta$ decay and the photon(s)
from the decay of the final nucleus excited state, might make up for the longer lifetime.
In order to find
a candidate for this final excited state, higher spins have to be
discarded since these decays are disfavored not only by the phase
space but also by the transition operator, which in order to couple
different spin states is necessarily of higher order in the current
---and hence much smaller in magnitude. However, excited $0_{1}^{+}$
states could have a chance, as they share the transition operator with the
decays to the ground state. In this new case the phase factor would disfavour
the decay but, if the NME pushed in the opposite direction and increased
it enough, then the lifetime of the process would not be much larger
than the corresponding to the $0_{g.s.}^{+}$. To explore this possibility,
we have computed, for the first time within the ISM, the NME's of the
$0\nu\beta\beta$ decays to excited $0_{1}^{+}$ states. 

\begin{table}

\caption{\label{cap:excited}NME's for the different decays to the different excited $0_{1}^{+}$ states studied.
The ground states are denoted by $gs$. Half-lives are calculated for
$\left\langle m_{\nu}\right\rangle =1$ eV. The minor differences between these NME's
and those of ref. \protect\cite{Caurier:2007wq} come as in fig. \protect\ref{cap:src}.}

\begin{center}
\smallskip
\begin{tabular*}{\linewidth}{@{\extracolsep{\fill}}c|cccccc}
\hline \noalign{\smallskip}

&
$A=48$&
$A=76$&
$A=82$&
$A=124$&
$A=130$&
$A=136$\tabularnewline
\hline
$M_{gs\rightarrow gs}^{0\nu\beta\beta}$&
0.61&
2.30&
2.18&
2.10&
2.12&
1.77\tabularnewline
$M^{0\nu\beta\beta}_{gs\rightarrow 0_{1}^{+}}$&
0.68&
1.49&
0.28&
0.80&
0.19&
0.49\tabularnewline
$\frac{G_{01}\left(gs\rightarrow gs\right)}{G_{01}\left(gs\rightarrow0_{1}^{+}\right)}$&
85&
12&
11&
40&
38&
22\tabularnewline
$T_{1/2}^{gs\rightarrow gs}$ ($10^{24}$ y)&
10.8&
7.70 &
1.94&
2.13&
1.29&
1.78\tabularnewline
$T_{1/2}^{gs\rightarrow0_{1}^{+}}$ ($10^{26}$ y)&
7.35 &
2.28&
12.9&
5.82&
61.2&
5.00\tabularnewline
\hline
\end{tabular*}\end{center}
\end{table}

The results are gathered in Table \ref{cap:excited}. With the only
exception of the relatively small increase of our result for $A=48$,
not really significant because of its huge space factor suppression,
we see that a common feature of all calculations is that the NME's for
the decays to $0_{1}^{+}$ states are smaller than the decays to the
ground state. This has been the case also in previous QRPA calculations. However, since these
results have been superseeded by more recent ones for the initial and final ground states 
---the case of refs. \cite{Suhonen:2000my,Suhonen:2000dr,Suhonen:2002wi,Suhonen:2003gx}, 
which do not consider the HOC relevant terms--- or do not treat SRC 
properly ---as happens in ref. \cite{Simkovic:2001qf}---,
new calculations for the transitions to final  $0_{1}^{+}$ states are required to
be comparable to our numbers. 

Table \ref{cap:excited} also includes the predicted half-lives for the transitions. 
We see that our results are typically two orders of magnitude longer
for the decays to excited states. The 
least disfavored $gs\rightarrow0_{1}^{+}$ transition would be that
of $^{76}$Ge, which is hindered by a factor $2.4$ from the NME times
$12$, the reduction factor coming from the phase space. This is,
in that case the $gs\rightarrow0_{1}^{+}$ transition is suppressed
by a factor $25-30$ compared to the $gs\rightarrow gs$, which is
probably too large to be compensated by the experimental gain via
background reduction. Nevertheless, it corresponds to experimentalists
to evaluate the practical interest of the decay to the excited $0^{+}$ in view of such a suppressed rate.

\section{Summary and Conclusions}

In this work we have calculated within the ISM framework the NME for
$0\nu\beta\beta$ decays of the emitters $^{48}$Ca, $^{76}$Ge, $^{82}$Se,
$^{124}$Sn, $^{128}$Te, $^{130}$Te and $^{136}$Xe. Also, we have studied the relative importance of different
contributions to the NME, namely HOC, FNS and SRC, concluding
that all of there is a nice agreement between the ISM and
different QRPA methods. However, the values of the  QRPA differ from the ISM ones when
the high seniority components are important in the latter.
We surmise that this difference may be due to the QRPA underestimation of the high seniority
components of the wave functions in these cases.
In addition, some discrepancies remain regarding the
importance of the effect of the Tensor part of the NME, and there is still
uncertainties in the treatment of the SRC's.

We have also studied the radial behaviour of the NME's, finding again agreement
between the ISM and the QRPA results.

Finally, we have calculated the NME's for the decay to excited $0_{1}^{+}$ states.
These transitions are suppressed with respect to the usual ones to
the ground state to a different extent for each decay, but
at least in a factor $25-30$ for the least hindered $A=76$.

\begin{ack}
This work has been supported by a grant of the Spanish Ministry of
Education and Science, FPA2007-66069, by the IN2P3-CICyT collaboration agreements,
by the Comunidad de Madrid (Spain), project HEPHACOS P-ESP-00346,
by the EU program  ILIAS N6 ENTApP WP1 and by the Spanish Consolider-Ingenio 2010 Program, CPAN (CSD2007-00042).

\end{ack}


\begin{thebibliography}{10}

\bibitem{PhysRevLett.81.1562}
Y. Fukuda et~al.,
\newblock Phys. Rev. Lett. 81 (1998) 1562.

\bibitem{PhysRevLett.89.011301}
Q.R. Ahmad et~al.,
\newblock Phys. Rev. Lett. 89 (2002) 011301.

\bibitem{PhysRevLett.90.021802}
K. Eguchi et~al.,
\newblock Phys. Rev. Lett. 90 (2003) 021802.

\bibitem{KlapdorKleingrothaus:2001ke}
H.V. Klapdor-Kleingrothaus, A. Dietz, H.L. Harney and I.V. Krivosheina,
\newblock Mod. Phys. Lett. A16 (2001) 2409.

\bibitem{KlapdorKleingrothaus:2004wj}
H.V. Klapdor-Kleingrothaus, I.V. Krivosheina, A. Dietz and O. Chkvorets,
\newblock Phys. Lett. B586 (2004) 198.

\bibitem{arnold:182302}
R. Arnold et~al.,
\newblock Phys. Rev. Lett. 95 (2005) 182302.

\bibitem{arnaboldi:142501}
C. Arnaboldi et~al.,
\newblock Phys. Rev. Lett. 95 (2005) 142501.

\bibitem{bloxham:025501}
T. Bloxham et~al.,
\newblock Phys. Rev. C76 (2007) 025501.

\bibitem{Avignone:2007fu}
F.T. Avignone~III, S.R. Elliott and J. Engel,
\newblock Rev. Mod. Phys. 80 (2008) 481.

\bibitem{Suhonen:1998ck}
J. Suhonen and O. Civitarese,
\newblock Phys. Rept. 300 (1998) 123.

\bibitem{Rodin:2006yk}
V.A. Rodin, A. Faessler, F. Simkovic and P. Vogel,
\newblock Nucl. Phys. A766 (2006) 107.

\bibitem{Caurier:2004gf}
E. Caurier et~al.,
\newblock Rev. Mod. Phys. 77 (2005) 427.

\bibitem{Retamosa:1995xt}
J. Retamosa, E. Caurier and F. Nowacki,
\newblock Phys. Rev. C51 (1995) 371.

\bibitem{Caurier:prl.77.1954}
E. Caurier, F. Nowacki, A. Poves and J. Retamosa,
\newblock Phys. Rev. Lett. 77 (1996) 1954.

\bibitem{Simkovic:1999re}
F. Simkovic, G. Pantis, J.D. Vergados and A. Faessler,
\newblock Phys. Rev. C60 (1999) 055502.

\bibitem{Kortelainen:2007rh}
M. Kortelainen and J. Suhonen,
\newblock Phys. Rev. C75 (2007) 051303.

\bibitem{Kortelainen:2007mn}
M. Kortelainen and J. Suhonen,
\newblock Phys. Rev. C76 (2007) 024315.

\bibitem{Caurier:2007wq}
E. Caurier, J. Men\'{e}ndez, F. Nowacki and A. Poves,
\newblock Phys. Rev. Lett. 100 (2008) 052503.

\bibitem{Feldmeier:1997zh}
H. Feldmeier, T. Neff, R. Roth and J. Schnack,
\newblock Nucl. Phys. A632 (1998) 61.

\bibitem{Simkovic:2007vu}
F. Simkovic et~al.,
\newblock Phys. Rev. C77 (2008) 045503.

\bibitem{Suhonen:2000my}
J. Suhonen,
\newblock Phys. Lett. B477 (2000) 99.

\bibitem{Suhonen:2000dr}
J. Suhonen,
\newblock Phys. Rev. C62 (2000) 042501.

\bibitem{Suhonen:2002wi}
J. Suhonen,
\newblock Nucl. Phys. A700 (2002) 649.

\bibitem{Suhonen:2003gx}
J. Suhonen and M. Aunola,
\newblock Nucl. Phys. A723 (2003) 271.

\bibitem{Simkovic:2001qf}
F. Simkovic et~al.,
\newblock Phys. Rev. C64 (2001) 035501.

\bibitem{Rodin:2007fz}
V.A. Rodin, A. Faessler, F. Simkovic and P. Vogel,
\newblock Nucl. Phys. A793 (2007) 213.

\bibitem{Muto:1994hi}
K. Muto,
\newblock Nucl. Phys. A577 (1994) 415c.

\bibitem{Doi:1980ze}
M. Doi et~al.,
\newblock Phys. Lett. B103 (1981) 219.

\bibitem{Doi:1985dx}
M. Doi, T. Kotani and E. Takasugi,
\newblock Prog. Theor. Phys. Suppl. 83 (1985) 1.

\bibitem{Suhonen:2008ijmpa}
J. {Suhonen} and M. {Kortelainen},
\newblock Int. J. Mod. Phys. E17 (2008) 1.

\bibitem{Wu:1985xy}
H.F. Wu et~al.,
\newblock Phys. Lett. B162 (1985) 227.

\bibitem{Miller:1975hu}
G.A. Miller and J.E. Spencer,
\newblock Ann. Phys. 100 (1976) 562.

\bibitem{Gniady:0000aa}
A. Gniady, E. Caurier and F. Nowacki,
\newblock to be published.

\bibitem{Caurier:2007qn}
E. Caurier, F. Nowacki and A. Poves,
\newblock Eur. Phys. J. A36 (2008) 195.

\bibitem{Kortelainen:2007rn}
M. Kortelainen, O. Civitarese, J. Suhonen and J. Toivanen,
\newblock Phys. Lett. B647 (2007) 128.

\bibitem{Roth:2005pd}
R. Roth et~al.,
\newblock Phys. Rev. C72 (2005) 034002.

\end{thebibliography}
\end{document}